\documentclass[twocolumn,aps, superscriptaddress]{revtex4}
\usepackage{amsmath}
\usepackage{amssymb}
\usepackage{graphicx}
\usepackage{color}
\usepackage{tabularx}

\newcommand{\be}{\begin{equation}}
\newcommand{\ee}{\end{equation}}
\newcommand{\bea}{\begin{eqnarray}}
\newcommand{\eea}{\end{eqnarray}}
\newcommand{\bse}{\begin{subequations}}
\newcommand{\ese}{\end{subequations}}
\newcommand{\parent}{EuMg$_2$Bi$_2$}
\newcommand{\Tn}{$T_{\textrm N}$}

\begin{document}

\author{Santanu Pakhira}
\affiliation{Ames Laboratory, Iowa State University, Ames, Iowa 50011, USA}
\author{Thomas Heitmann}
\affiliation{The Missouri Research Reactor and Department of Physics and Astronomy, University of Missouri, Columbia, Missouri 65211, USA}
\author{S. X. M. Riberolles}
\affiliation{Ames Laboratory, Iowa State University, Ames, Iowa 50011, USA}
\author{\mbox{B. G. Ueland}}
\affiliation{Ames Laboratory, Iowa State University, Ames, Iowa 50011, USA}
\author{R. J. McQueeney}
\author{D. C. Johnston}
\author{David Vaknin}
\affiliation{Ames Laboratory, Iowa State University, Ames, Iowa 50011, USA}
\affiliation{Department of Physics and Astronomy, Iowa State University, Ames, Iowa 50011, USA}

\title{Zero-field magnetic ground state of \parent }
\date{\today}

\begin{abstract}
Layered trigonal \parent\ is reported to be a topological semimetal that hosts multiple Dirac points that may be gapped or split by the onset of magnetic order.  Here, we report zero-field single-crystal neutron-diffraction and bulk magnetic susceptibility measurements versus temperature $\chi(T)$ of {\parent} that show the intraplane ordering is ferromagnetic (Eu$^{2+},\, S= 7/2$) with the moments aligned in the $ab$-plane while  adjacent layers are aligned antiferromagnetically  (i.e., A-type antiferromagnetism) below the N\'eel temperature.   
\end{abstract}

\maketitle

\section{Introduction}

Recent studies of rare-earth-based metallic systems have revealed novel electronic states arising from a complex interplay of magnetism and electron-band topology~\cite{hirschberger2016chiral,shekhar2018anomalous,borisenko2019time,soh2019magnetic,wang2016anisotropic,jo2020manipulating,riberolles2020magnetic}. EuMg$_2$Bi$_2$ is one such system that undergoes antiferromagnetic (AFM) ordering below a N\'eel temperature $T_{\rm N} \approx 6.7$~K~\cite{Pakhira2020,may2011structure, kabir2019observation} and is also reported to host multiple Dirac points located at different energies with respect to the Fermi energy~\cite{kabir2019observation}.  Various topological states of \parent\ (such as axion  or Weyl states) are dependent on the nature of the magnetic order since time-reversal symmetry breaking and magnetic crystalline symmetry may gap or split the Dirac points.

\parent\ crystallizes in the trigonal CaAl$_2$Si$_2$-type crystal structure~\cite{zheng1986site} (space group $P\bar{3}m1$, No. 164), where the Eu atoms form a triangular lattice in the  $ab$~plane with simple hexagonal-stacking along the $c$~axis. Recently, our  anisotropic magnetic susceptibility $\chi(T)$  data measured in a magnetic field $H= 1$~kOe demonstrated that both the in-plane and out-of-plane magnetic susceptibilities are almost temperature independent below $T_{\rm N}$~\cite{Pakhira2020}. Using our recent formulation of molecular-field theory~\cite{johnston2012magnetic,johnston2015unified}, it has been proposed that the magnetic structure below $T_{\rm N}$ is a $c$-axis helix with a turn angle of $\approx 120^\circ$ between adjacent Eu layers in which the Eu spins are ferromagnetically aligned in the $ab$~plane in each Eu layer~\cite{Pakhira2020}.

Here, we report neutron-diffraction measurements on single-crystal EuMg$_2$Bi$_2$  and determine the zero-field Eu$^{2+}$ spin $S=7/2$ magnetic structure below $T_{\rm N}$ to be \mbox{A-type} AFM order with the moments aligned in the $ab$~plane.   We also present $\chi(T)$ results  in a low magnetic field $H = 100$~Oe that are consistent with the magnetic structure obtained from neutron diffraction measurements in zero field.  The difference between the present AFM structure and that inferred from the previous $\chi(T)$ measurements in $H = 1$~kOe which report a 120 degree helical structure \cite{Pakhira2020} implies that the magnetic texture (i.e., structure and/or domains) is sensitive to the strength of the applied magnetic field and requires additional neutron-diffraction measurements under magnetic field for confirmation.

The experimental details and methods are presented in Sec.~\ref{Sec:ExpDet}.  The neutron diffraction measurements and analyses are discussed in Sec.~\ref{Sec:Neutron} and the $\chi(T)$ measurements in Sec.~\ref{Sec:MagSus}.  The  results are summarized in  Sec.~\ref{Sec:Conclu}.

\section{\label{Sec:ExpDet} Experimental Details and Methods}

\parent\ single crystals with hexagonal lattice parameters $a = 4.7724(3)$ and $c = 7.8483(5)$~\AA~\cite{Pakhira2020} were grown by a self-flux method with starting composition EuMg$_4$Bi$_6$ as described previously~\cite{may2011structure}. The $\chi(T)$ measurements were carried out using a magnetic-properties measurement system (MPMS, Quantum Design, Inc.) in the temperature range 1.8--300~K\@. A $\sim 50$~mg crystal was cut into two pieces having masses $\sim 10$~mg and $\sim 40$~mg. The 10~mg piece was used for the magnetization measurements and the 40~mg piece was used for neutron diffraction experiments.

Single-crystal neutron-diffraction experiments were performed in zero applied magnetic field using the TRIAX triple-axis spectrometer at the University of Missouri Research Reactor (MURR). An incident neutron beam of energy $E_i = 30.5$ meV or 14.7 meV was directed at the sample using a pyrolytic graphite (PG) monochromator. Elastic scattering data were acquired with $E_i = 30.5$ meV in order to reduce the absorption caused by highly absorbing Eu, whereas $E_i = 14.7$~meV was used to improve the resolution in a search for possible peaks associated with an incommensurate magnetic structure. A PG analyzer was used to reduce the background. Neutron wavelength harmonics were removed from the beam using PG filters placed before the monochromator and in between the sample and analyzer. Beam divergence was limited using collimators before the monochromator; between the monochromator and sample; sample and analyzer; and analyzer and detector of $60^\prime-60^\prime-40^\prime-40^\prime$, respectively.

A 40~mg \parent\ crystal was mounted on the cold tip of an Advanced Research Systems closed-cycle refrigerator with a base temperature of 4~K\@. The crystal was aligned in the $(HHL)$ and $(H0L)$ scattering planes whereupon a wide range of reciprocal space was accessible for our comparative diffraction study above (10~K) and below (4~K) $T_{\rm N} = 6.7$~K\@. Reciprocal space was searched extensively using a series of $H$-, $HH$-,  and $L$-scans as well as mesh scans in order to identify any commensurate or incommensurate wave vectors that might be present.

\section{\label{Sec:Neutron} Neutron diffraction}

\begin{figure}
\centering
\includegraphics[width=3.4 in]{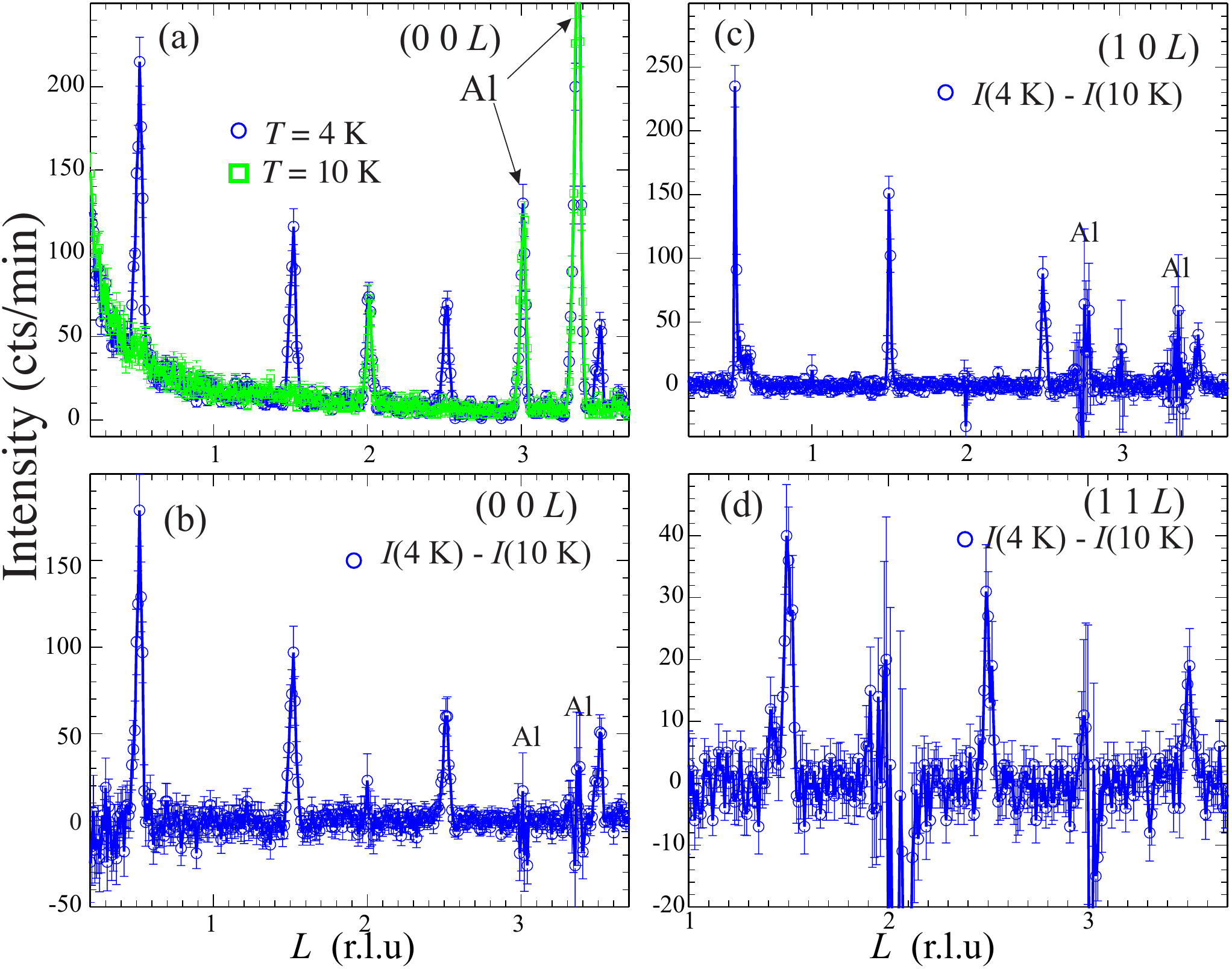}
\caption{(a) Diffraction pattern along $(00L)$ of single-crystal \parent\ at 4 and 10 K as indicated.  Aluminum Bragg reflections  are from the sample holder. (b) Difference between the $(00L)$ patterns taken at 4 K and 10 K\@. (c) Difference between the $(10L)$ patterns taken at 4 K and at 10 K\@. (d) Difference between a $(11L)$ patterns taken at $T=4$~K and 10 K\@. All three difference patterns show clear magnetic peaks at half-integer $L$ up to $L=3.5$, consistent with A-type AFM, i.e, the $H = 0$ ground state is such that the intraplane ordering is ferromagnetic while adjacent layers are aligned antiferromagnetically.}
\label{Fig:00l}
\end{figure}

\begin{figure}
\centering
\includegraphics[width=3.4 in]{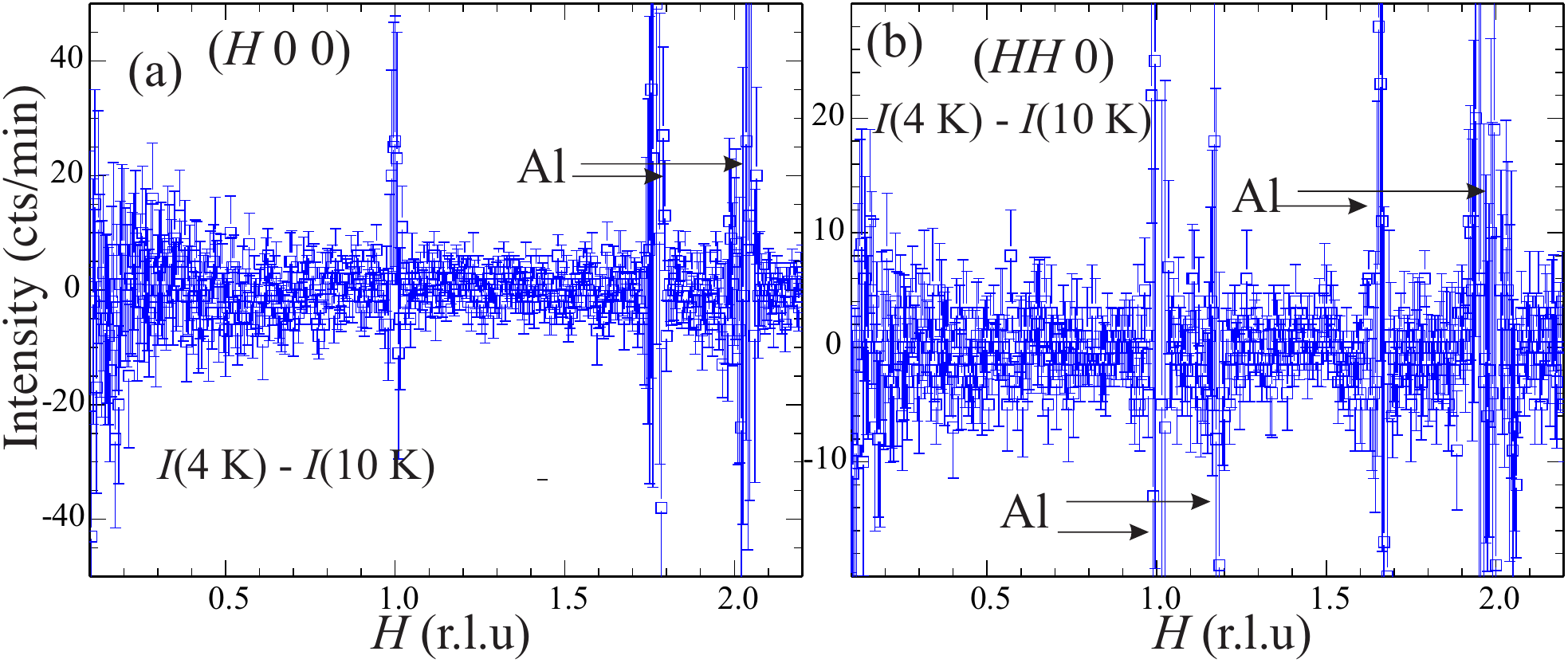}
\caption{Difference between the pattern taken at $T = 4$~K and that at 10 K for (a)  $(H 0 0)$ and for (b) $(H H 0)$ with no indication of nonferromagnetic in-plane magnetic ordering of \parent, that together with Fig.\ \ref{Fig:00l} indicate simple A-type antiferromagnetism. The noise in (b) is due to thermal changes of the Bragg peaks, most prominent are those from the Al can containing the sample.}
\label{Fig:hh0}
\end{figure}

Figure~\ref{Fig:00l}(a) shows diffraction scans along $(00L)$  at 4 and 10 K, where reflections at half-integer $L$ values are apparent at $T = 4$~K\@.  For more clarity, Fig.~\ref{Fig:00l}(b) shows the difference between these two scans, where  within experimental uncertainty there is no evidence for other reflections associated with a modulated structure along the $c$ axis.  Similar differences [i.e., $I$(4 K) $-$ $I$(10 K)] for scans along $(10L)$ and $(11L)$, shown in Figs.~\ref{Fig:00l}(c) and~\ref{Fig:00l}(d), respectively, also reveal new peaks at half-integer $L$ values. Qualitatively, these newly emerging Bragg reflections indicate the doubling of the unit cell along the $c$~axis. We also note that the intensities of the new peaks become weaker at larger $L$ values and also as the total momentum transfer $Q$ gets larger  [i.e., $Q_{(11L)} >  Q_{(10L)}> Q_{(00L)}$], roughly following the falloff expected from the magnetic form factor. These qualitative observations unequivocally establish that these peaks are associated with A-type AFM ordering with AFM propagation vector $\vec{\tau} = \left(0,0,\frac{1}{2}\right)$ (in reciprocal-lattice units) consisting of ferromagnetic layers with moments aligned in the $ab$~plane that are stacked antiferromagnetically.  The $\chi(T)$ data in the following section confirm that the ordered moments lie in the $ab$~plane.

To confirm the in-plane ferromagnetic (FM) structure we carried out more comprehensive scans to search for additional magnetic peaks. In particular, Fig.~\ref{Fig:hh0} shows that no additional magnetic peaks are observed in the difference between scans taken at 4 and 10~K along $(H00)$ (a)  and $(HH0)$ (b), consistent with a single AFM propagation vector $\vec{\tau} = \left(0,0,\frac{1}{2}\right)$.  The sharp features in these difference scans are artifacts of the subtraction caused by slight shifts in nuclear Bragg-peak positions due to thermal expansion upon heating. We also performed scans in the $(HHL)$ and $(H0L)$ planes and found additional peaks only at the expected half-integer $L$~positions.

\begin{figure}
\centering
\includegraphics[width=2.75 in]{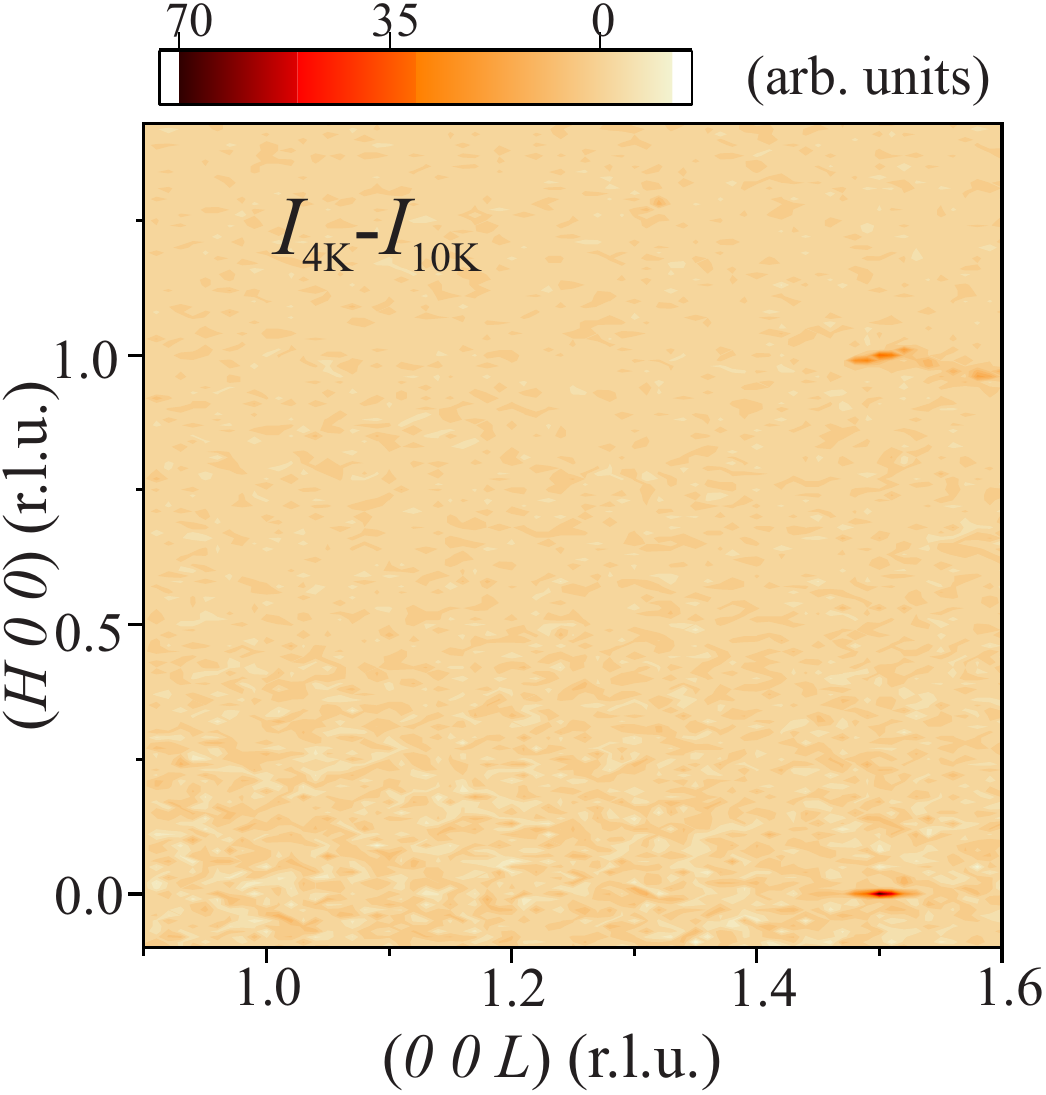}
\caption{2D $(00L)$ $(H00)$ mesh at $E_i=14.4$ meV measured at 4 K and after subtracting a similar mesh at 10 K, i.e., in the paramagnetic state above $T_{\rm N}$. The reflections (0 0 1.5) and (1 0 1.5) are purely magnetic peaks due to a 180$^\circ$ rotation between adjacent layers. The absence of other features in the mesh constitute evidence that there is no 120$^\circ$ helical order at zero applied magnetic field.}
\label{Fig:Mesh}
\end{figure}

A mean-field analysis of previous single-crystal $\chi(T)$ measurements with \mbox{$H = 1$~kOe} (as opposed to the zero applied magnetic field for the present neutron-diffraction experiments) indicated a $c$-axis helical magnetic ground state where each adjacent Eu-moment layer  is ferromagnetically-aligned in the $ab$~plane and rotated by $\approx 120^\circ$ with respect to its nearest-neighbor (NN) Eu layers~\cite{Pakhira2020}. If present, such a magnetic structure would give rise to a magnetic unit cell three times that of the  chemical unit cell along the $c$ axis, and would be manifested by extra magnetic Bragg reflections shifted from the nuclear Bragg positions by $\pm 1/3$. To search for such reflections or other helical magnetic structures, we conducted  scans  around prominent magnetic peaks using  $E_i = 14.7$ meV.  Figure~\ref{Fig:Mesh} shows a $(H0L)$ 2D map of the intensity at $T = 4$~K minus that taken at 10~K\@.  As shown, we only find peaks at (0~0~1.5) and at (1~0~1.5) associated with A-type AFM order and observe no other features, in particular no peaks are found at $L \pm 1/3$ that would correspond to the 120$^\circ$ rotation between NN layers.  Nevertheless, we note that the magnetic Bragg reflections are elongated along the $(00L)$ direction beyond the instrumental resolution. Such a shape may arise from stacking faults of the FM layers.

\begin{figure}
\includegraphics[width=3. in]{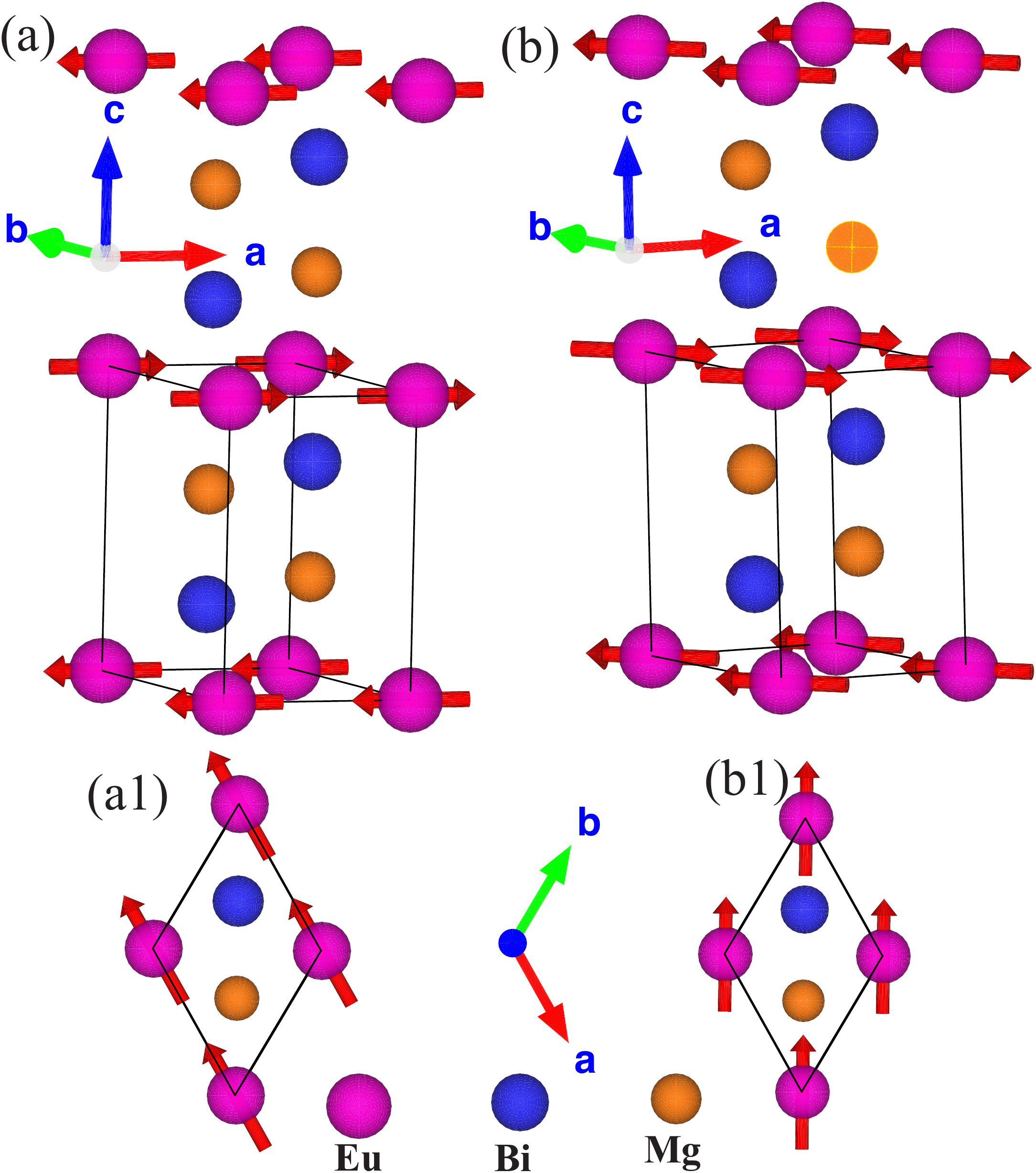}
\caption{Chemical and A-type AFM spin structure  of \parent.  (a) FM spins are aligned  towards NN and (b) towards NNN. (a1) and (b1) show the corresponding projection of a single layer on the $ab$-plane. Our neutron diffraction data are insensitive to the direction of the FM moment in the plane.}
\label{Fig:structure}
\end{figure}

The proposed A-type AFM structure is shown in Fig.~\ref{Fig:structure}, where adjacent NN FM layers are rotated by 180$^\circ$ with respect to each other.   The direction of the FM moment in the Eu layer cannot be determined from neutron diffraction alone.  Thus, in Fig.~\ref{Fig:structure}  we show two possible magnetic structures where the moments are pointing towards their in-plane NN (a,a1) or to their next nearest neighbor (NNN) (b,b1) (there are no additional possibilities according to the Bilbao crystallographic server~\cite{Mato2015}).  Using published values for the structural parameters, we obtain good agreement with the intensities of the nuclear Bragg peaks, both above and below TN.  From this basis, we are able to confirm the A-type magnetic structure and obtain an estimate for the ordered magnetic moment $\mu = \langle gS\rangle\,\mu_{\rm B}$ at $T = 4$ K using the FullProf software~\cite{RODRIGUEZCARVAJAL1993}.

Individual Bragg peaks measured by $\theta$-$2\theta$ scans were fit to Gaussian lineshapes to determine  their integrated intensities which were then corrected for the geometric Lorentz factor.  To account for the significant  neutron absorption cross section of Eu, we use the  \textsc{mag2pol} \cite{Qureshi2019} software, by supplying the approximate sample shape as a plate of  dimensions $2\times2\times0.5$ mm$^3$.  For the refinement of the chemical structure with space group $P{\bar 3}m1$ we used published  structural parameters~\cite{may2011structure,Pakhira2020} which we find are in good agreement  with our refinement.  As noted above, the possible magnetic structures that can occur with a second order phase transition from space group $P{\bar 3}m1$ to AFM order with propagation vector $\vec{\tau}= \left(0, 0, \frac{1}{2}\right)$ are consistent with antiparallel   $c$~axis  stacking of FM layers (A-type AFM order]. In our analysis of the magnetic structure, we use  the $C_c2/m$ (\# 12.63) symmetry~\cite{Gallego2019} [this is the magnetic structure shown in Fig.\ \ref{Fig:structure}(a) with magnetic moments directed towards NN), and note that our diffraction data eliminates any other minimum symmetry reduction. 

\begin{figure}
\centering
\includegraphics[width=3. in]{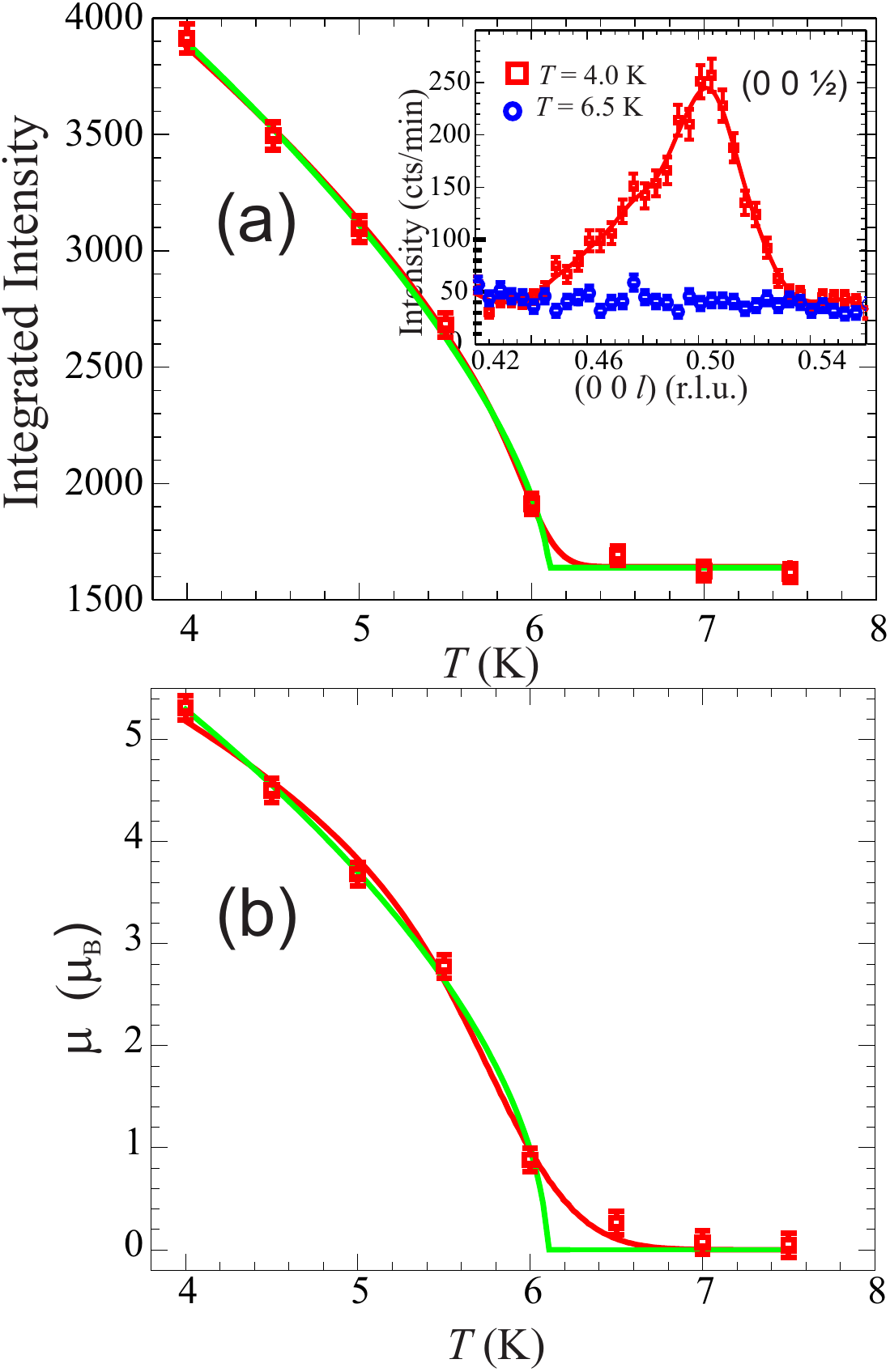}
\caption{(a) Integrated intensity as a function of temperature $T$ of the (0 0 $\frac{1}{2}$) magnetic Bragg reflection and (b)~calculated ordered moment~$\mu$ $versus$ $T$, with a power-law fit (solid green line) indicating $T_{\rm N} = (6.2 \pm 0.4)$~K\@.  The red curves in (a) and~(b) assume a Gaussian dstribution of $T_{\rm N}$.}
\label{Fig:OP}
\end{figure}

Our refinement of the magnetic structure also yields an average magnetic moment $\mu = \langle gS\rangle\,\mu_{\rm B} = (5.3\pm0.5$)\,$\mu_{\rm B}$ at $T = 4$~K. This value is smaller than the zero-temperature ordered moment $\mu = 7\,\mu_{\rm B}$ expected from the electronic configuration of Eu$^{2+}$~\cite{Cable1977} with $S=7/2,\ L = 0$ and $g=2$ because  $\mu$  is not yet saturated to its full value at $T = 0$.  Figure~\ref{Fig:OP}(a) shows the integrated intensity of the (0 0 0.5) magnetic peak as a function temperature where we use a simple power-law function $I_{\rm (0\,0\, 0.5)}(T) = C|1-T/T_{\rm N}|^{2\beta} \propto \mu^2$ to fit the data (solid line with sharp transition).  The smooth line around {\Tn} is obtained by the same power law but weighted by a Gaussian distribution of {\Tn} (this form is sometimes used to account for crystal inhomogeneities) yielding $T_{\rm N} = 6.2 \pm 0.4$ and $\beta = 0.40 \pm 0.05$. The temperature probe in the neutron diffraction measurements is placed outside the helium-filled aluminum can holding the crystal, likely recording temperatures that are slightly lower than that of the sample. This may explain the discrepancy with the {\Tn}\ measured by  the magnetic  susceptibility.   Most importantly, the phenomenological fits show that the order parameter is still increasing at $T = 4$~K and not close to its saturated value.  Indeed, Fig.~\ref{Fig:OP}(b) shows the square root of the data in Fig.~\ref{Fig:OP}(a) after subtracting the background and normalizing the value at $T = 4$ K to the extracted average magnetic moment to $\mu(4~ {\rm K}) = 5.3$ $\mu_{\rm B}$. Using the power-law  yields ($9.5 \pm 1$) $\mu_{\rm B}$ at $T=0$.  This approach overestimates the expected 7 $\mu_{\rm B}$ at $T=0$ because the phenomenological  power-law fit is only accurate just below $T_{\rm N}$~\cite{johnston2015unified}.

\section{\label{Sec:MagSus} Magnetic Susceptibility}

\begin{figure}[ht!]
\includegraphics[width=3. in]{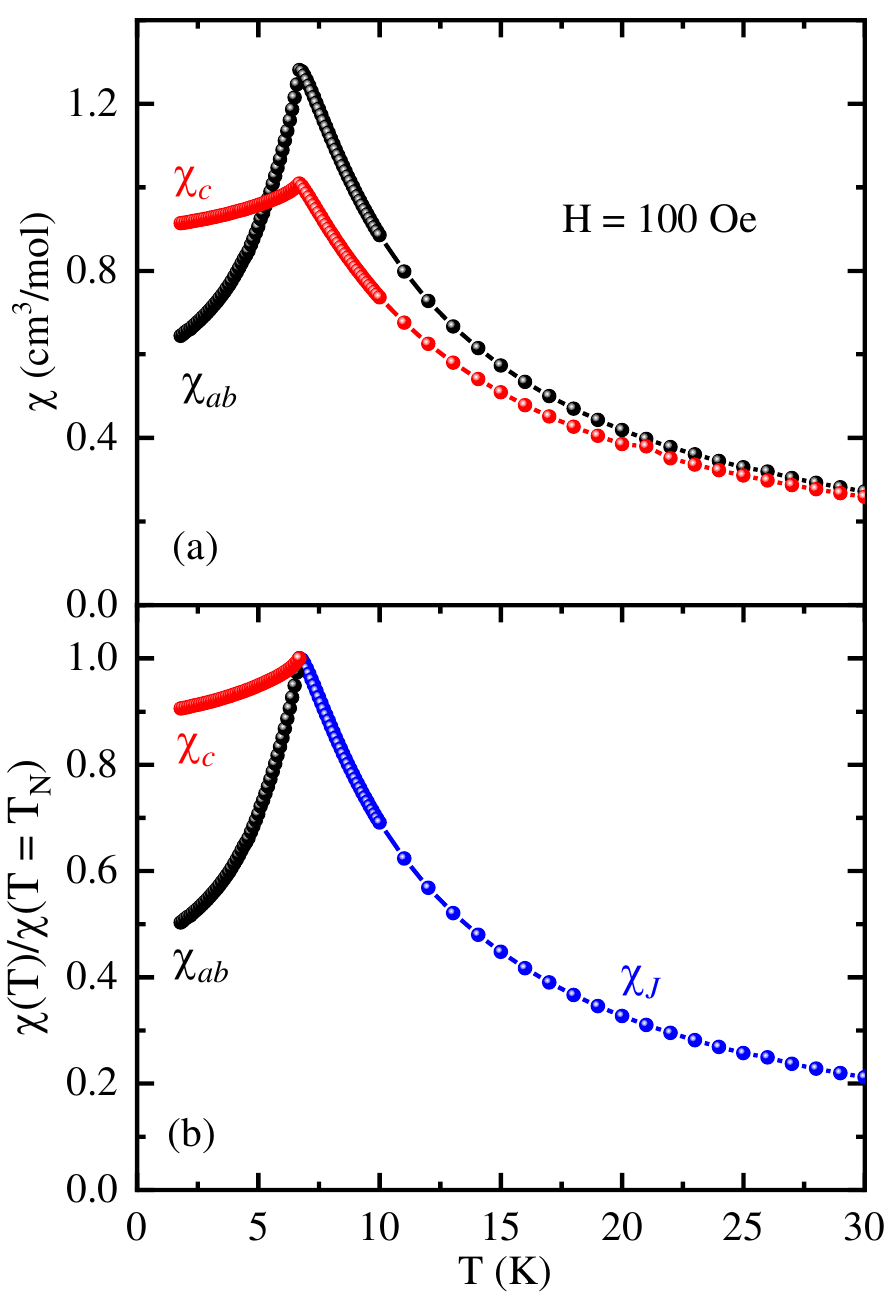}
\caption{Temperature dependence of zero-field-cooled (ZFC) magnetic susceptibility measured at different applied magnetic fields as listed when the field is (a) in the $ab$-plane ($H \parallel ab$) and (b) along the $c$-axis ($H \parallel c$). In panel~(b), the two data sets for $H \parallel ab$ and $H \parallel c$ perfectly overlap in the paramagnetic regime with $T \geq T_{\rm N}$\@.}
\label{Fig:chi}
\end{figure}

We presume that the difference between zero-field A-type AF order determined from neutron diffraction and the reported 120$^{\circ}$  helical order inferred  from bulk susceptibility is caused by the application of a magnetic field.~\cite{Pakhira2020}.  Accordingly, we measured the temperature dependence of the magnetic susceptibility $\chi \equiv \frac{M}{H}$ a very low-field of at $H = 100$~Oe   to better approximate the zero-field conditions of our neutron diffraction data (on the same piece of single crystal) as shown in Figs.~\ref{Fig:chi}(a) and \ref{Fig:chi}(b) for $H$ aligned in the $ab$~plane ($\chi_{ab}$) and along the $c$~axis ($\chi_c$), respectively.  The compound orders antiferromagnetically below the N\'eel temperature $T_{\rm N} \approx 6.7$~K, as reported earlier~\cite{Pakhira2020, may2011structure, kabir2019observation}. Although $\chi_c$ in Fig.~\ref{Fig:chi}(a) is nearly independent of $T$ below $T_{\rm N}$, $\chi_{ab}(H=100$~Oe) in Fig.~\ref{Fig:chi}(a) decreases by about a factor of two upon cooling from $T_{\rm N}$ to 1.8~K\@.

To clarify the nature of the ground-state magnetic structure, we analyzed the low-field $\chi(T)$ data in Fig.~\ref{Fig:chi}(a) using unified molecular-field theory (MFT)~\cite{johnston2012magnetic,johnston2015unified}. This theory holds for systems of identical crystallographically-equivalent Heisenberg spins interacting by Heisenberg exchange and the magnetic properties are calculated from the exchange interactions between an arbitrary spin and its neighbors. According to the MFT, for a $c$-axis helix $\chi_c$ is independent of $T$ below $T_{\rm N}$, as seen to be approximately satisfied in Fig.~\ref{Fig:chi}(a). However, $\chi_{ab}$  is dependent on  the turn angle $kd$ for a $c$-axis helix and are related by
\bea
\ \frac{\chi_{Jab} (T = 0)}{\chi_J (T_{\rm N})} = \frac{1}{2[1 + 2\cos(kd) + 2\cos^2(kd)]},
\label{Eq.Turnangle}
\eea
where $k$ is the magnitude of the $c$-axis helix wave vector in reciprocal-lattice units, $d$ is the distance between the magnetic layers along the $c$~axis, and the subscript $J$ represents that the anisotropy in $\chi (T \geq T_{\rm N})$ has been removed by spherically averaging the anisotroic \mbox{$\chi (T \geq T_{\rm N})$} data; hence the Heisenberg interactions~$J$ determine the resulting behavior of the spherically-averaged magnetic susceptibility above~$T_{\rm N}$\@.

Figure~\ref{Fig:chi}(b) depicts the normalized susceptibility $\chi(T)/\chi(T_{\rm N})$ of \parent\ for $H \parallel ab$ and $H \parallel c$, respectively, obtained from the data in Fig.~\ref{Fig:chi}(a). It is evident that $\chi_{ab}(1.8~{\rm K})/\chi(T_{\rm N}) \approx 0.5$, yielding a turn angle $kd \approx 180^\circ$ from Eq.~(\ref{Eq.Turnangle}).  This turn angle corresponds to A-type AFM order, in agreement with the above analysis of the neutron-diffraction measurements below $T_{\rm N}$ in zero applied field. The same value of $\chi_{ab}(1.8~{\rm K})/\chi(T_{\rm N}) \approx 0.5$ at $T=0$ is obtained from a calculation for equal populations of three collinear AFM domains oriented at 120$^\circ$ from each other.  We also note that good fits to $\chi_{ab}(T)$ data obtained in $H=1$~kOe for EuCo$_2$P$_2$ and EuNi$_{1.95}$As$_2$ crystals with the tetragonal ${\rm ThCr_2As_2}$ crystal structure were obtained for $c$-axis helical structures with turn angles in good agreement with the respective $c$-axis helical structures previously obtained from zero-field neutron-diffraction measurements~\cite{Sangeetha2016, Sangeetha2019}. 

\section{\label{Sec:Conclu} Conclusion}

EuMg$_2$Bi$_2$ has drawn interest as it exhibits  electronic topological properties that give rise to  Dirac-like bands near the Fermi level.  The presence of the large-spin element Eu$^{2+}$ in the compound makes it attractive since magnetic order can introduce a gap or lower the degeneracy of the Dirac-like bands to create more exotic states, for instance Weyl states.  Here, we use zero-field single-crystal zero-field neutron diffraction and low-field magnetic susceptibility measurements  to determine the magnetic ground state of this system.

The neutron-diffraction experiments reveal  that the intraplane ordering of Eu$^{2+}  (S= 7/2$) is ferromagnetic with $ab$-plane alignment and that  adjacent layers are  stacked antiferromagnetically  (i.e., A-type AFM order). Our detailed analysis also confirms that the ordered magnetic moment, as $T$ approaches 0 K, attains its expected  value $\sim 7\,\mu_{\rm B}$/Eu. The temperature-dependent magnetic susceptibility measurements at a very low magnetic field applied along the $c$-axis and in the $ab$-plane are consistent with  the A-type antiferromagnetism  below $T_{\rm N} = 6.7$~K and also that the moments  are aligned in the $ab$ plane. We note that close examination of the magnetic Bragg-reflection peak-shapes exhibit broadening along the $(0 0 L)$ direction indicating imperfect correlations between the antiparallel-stacked FM layers.  Previous $\chi(T)$ measurements in $H = 1$~kOe indicated that the magnetic structure is a $c$-axis helix with a $120^\circ$ turn angle instead of the A-type AFM structure (180$^\circ$ $c$-axis helix) obtained from our zero-field neutron-diffraction measurements.   Neutron-diffraction studies under applied magnetic fields are required to confirm the evolution of the magnetic structure with field inferred from our zero-field neutron-diffraction measurements and the 1~kOe magnetic-susceptibility measurements and are planned for the future.

\acknowledgments

This research was supported by the U.S. Department of Energy, Office of Basic Energy Sciences, Division of Materials Sciences and Engineering.  S.X.M.R.\ and B.U.\ are supported by the Center for Advancement of Topological Semimetals, an Energy Frontier Research Center funded by the U.S. Department of Energy Office of Science, Office of Basic Energy Sciences, through Ames Laboratory. Ames Laboratory is operated for the U.S. Department of Energy by Iowa State University under Contract No.~DE-AC02-07CH11358.


\end{document}